# EXTRINSIC SIZE EFFECT OF PYROELECTRIC RESPONSE OF FERROELECTRIC FILMS

*Eugene A. Eliseev[1], Nicholas V. Morozovsky[2*], Mykola Ye. Yelisieiev[3],*
*and Anna N. Morozovska[2†],*

[1]*Institute for Problems of Materials Science, National Academy of Sciences of Ukraine,*
*Krjijanovskogo 3, 03142 Kyiv, Ukraine*

[2] *Institute of Physics, National Academy of Sciences of Ukraine,*
*pr. Nauky 46, 03028 Kyiv, Ukraine*

[3] *Secondary School 108, Uralka str. 8, 03028 Kyiv, Ukraine*

**Abstract**

We developed a theoretical model for the calculation of the main characteristics of pyroelectric response of pyroactive multilayer structures, which are basic elements of modern pyroelectric detectors of radiation and other transducers.

In the framework of the proposed model, the dependences of the pyroelectric response frequency spectra on the ferroelectric film thickness have been investigated. It was found that a pronounced extrinsic size effect is inherent to the pyroelectric response frequency spectra, such as the strong thickness dependence of the response maximal value and its frequency position.

Thickness dependences of the frequency position of the pyroelectric response maximum and the optimal thickness of the pyroactive ferroelectric film are calculated. The frequency spectra of the pyroelectric current and voltage calculated theoretically are in quantitative agreement with experimental ones for Al/P(VDF-TrFE)/Ti/Si pyroactive structure.

Developed procedure is important for improving the sensitivity of pyroelectric devices. Obtained analytical expressions allow us to optimize the performances of pyroactive elements using the existence of the size effect, which depend on pyroactive layer and substrate thicknesses, as well as on the conditions of heat transfer and exchange at corresponding boundaries of the pyroactive structure.

---

[*] corresponding author, e-mail: nicholas.v.morozovsky@gmail.com

[†] corresponding author, e-mail: anna.n.morozovska@gmail.com

# I. INTRODUCTION

Pyroelectric effect in polar solids is a consequence of the polarization temperature dependence, and it leads to the appearance of uncompensated electric charges of opposite sign at certain crystallographic surfaces (polar cuts) under the temperature change. Based on the theory of symmetry, any material that unit cell has no inversion center can be pyroelectric [1, 2, 3].

Comprehensive studies of pyroelectric response features of ferroelectric materials and films had been started in the last century [1-3, 4, 5, 6, 7] and are actively continued nowadays owing to the multitude applications of pyroelectrics in sensorics and thermal imaging systems [7, 8, 9, 10, 11, 12].

From the very beginning the general principles of creation of pyroelectric detectors of radiation (PDR) have been established [13, 14] and the main details of registration methods and relevant equipment have been specified [14, 15, 16, 17, 18, 19]. Subsequently the calculations of PDR basic characteristics, including thermal analysis, were performed and the main types of PDR were realized [1 - 11, 16-19, 20, 21, 22, 23, 24, 25, 26, 27, 28, 29, 30, 31, 32, 33] (see also periodical guides on pyroelectricity by S. B. Lang [34, 35, 36]).

During the development of PDR technologies the transition from assembled structures [1, 2, 5] to hybrid and micro-machined ones [8-11, 37, 38] had taken place, and then it was followed by the transition to monolithic integrated structures [11, 12, 22-32, 39, 40, 41, 42, 43]. Note, that the problem of pyroactive structures integrating into Si-basis of micro- and nano-electronics could not be solved without the transition from the bulk to the thin film technologies. At that the bulk-type [1-5, 17-22] and membrane-type [37, 38] structures were replaced by multilayer film-on-substrate type structures [11, 22-33, 39-43]. As distinct from the bulk-type structures, operational effectiveness of the film-on-substrate type structures is defined not only by pyroelectric figures of merits of a ferroelectric film material [1-5], but also by thermal and geometrical parameters of the film and substrate [11, 23-33, 39]. For the film-on-substrate type structures, intensive thermal exchange between the film and substrate is inevitable in the definite operating frequency range of PDR. Owing to the heat sink into substrate a decrease of the film temperature increment takes place in the frequency range. This effect leads to the appreciable decrease of the film-on-substrate pyroelectric response in comparison with the response of freely suspended film [23-26, 28]. The maximum possible substrate thinning was proposed for the effect minimization in membrane-type structures [37, 38].

The addition of intermediate thermal isolating layer was proposed for thermal decoupling of the film and substrate [28-30, 38], and the effectiveness of the air gap was also considered [44, 45]. With the same purpose integrated structures with PVDF/TrFE copolymer [30, 32] and with $SiO_2$-aerogel [41-43] thermal isolating layer were studied and then fabricated. The structures with mesoscopic porous silicon interlayer of low thermal diffusivity have been studied theoretically and experimentally [46, 47, 48].

The fundamental relationship $\lambda_T = \sqrt{2a_T/\omega}$ between the penetration depth of the temperature wave $\lambda_T$ [27, 28, 49], named also the thermal diffusion length [33, 45], and modulation frequency of the thermal flux $\omega$ shows that the thermal decoupling of the film and substrate can be achieved by decreasing $\lambda_T$, i.e. by increasing $\omega$ ($a_T$ is the diffusivity of the media, where the temperature wave propagates). Therefore the question about the optimal configuration of the film-on-substrate type structure for operation at a given frequency (or in the certain frequency range) was raised and considered [22, 23, 28, 29].

Optimal thickness of pyroelectric layer $h_{opt}$ at a specified $\omega$ was determined by Holeman [23] for a specific type of the boundary conditions. Schopf et al [28] revealed the weak dependence of $h_{opt}$ on the substrate thickness under the practical equality of $h_{opt}$ and $\lambda_T$. The dependence of $h_{opt}$ on $\omega$ (and so on $\lambda_T$ value) was also pointed out by Mahrane et al [29]. Thus the existence of the optimal relationships between thermal and geometric parameters of the pyroactive layer, substrate and modulation frequency necessary to achieve the maximal value of pyroelectric response are very likely. The influence of the substrate thickness and relative effusivities of pyroelectric layer and substrate on pyroelectric response has been considered [23]. Also the influence of the thickness of substrate and its thermal parameters [23, 25, 38], or thermal quality of substrate [24], or the thickness of pyroelectric layer [29] on pyroelectric response was considered. To the best of our knowledge the optimal relationship has not been established yet, and that is why we decided to consider the problem in this work.

The above facts motivated this work aimed to identify and study size effects and frequency spectrum of pyroelectric voltage and current response of the multi-layer structures. The inhomogeneous temperature distribution is characteristic for the film-on-substrate type structures with unequal thermal parameters. Therefore we solved the heat conduction equations for each of the layer coupled by the boundary conditions at the interfaces. The temperatures are equal and heat fluxes are continuous at the layers' boundaries, and the thermal losses exist at the outer boundaries. The temperature distribution in the pyroactive layer was obtained as the solution of the heat problem. The temperature was substituted in the expressions for pyroelectric voltage under the open circuit conditions, or in the expressions for pyroelectric current under the short circuit conditions. Both these cases were considered in refs [3, 11, 39, 40]. Size effects are considered in the narrow sense as the dependence of pyroelectric response on the layers' thicknesses. Theoretical results are compared with experiment.

The paper is organized in the following way. The problem statement for the calculation of the multilayer pyroelectric response is presented in the Section II. Analytical solution for two-layer heat conduction and electric problems is analyzed in the Section III.A. The Section III.B includes the study of the extrinsic size effect and is followed by the comparison with experimental data (Section III.C). The Section IV is a conclusion.

## II. PYROELECTRIC RESPONSE OF A MULTILAYER SYSTEM

**Figures 1a-c** gives the schematic representation of realistic, simplified and modeled two-layer pyroactive structures. The thicknesses $h_1$ and $h_2$ of pyroactive ferroelectric film and substrate respectively and their coordinates are shown in **Figure 1c**

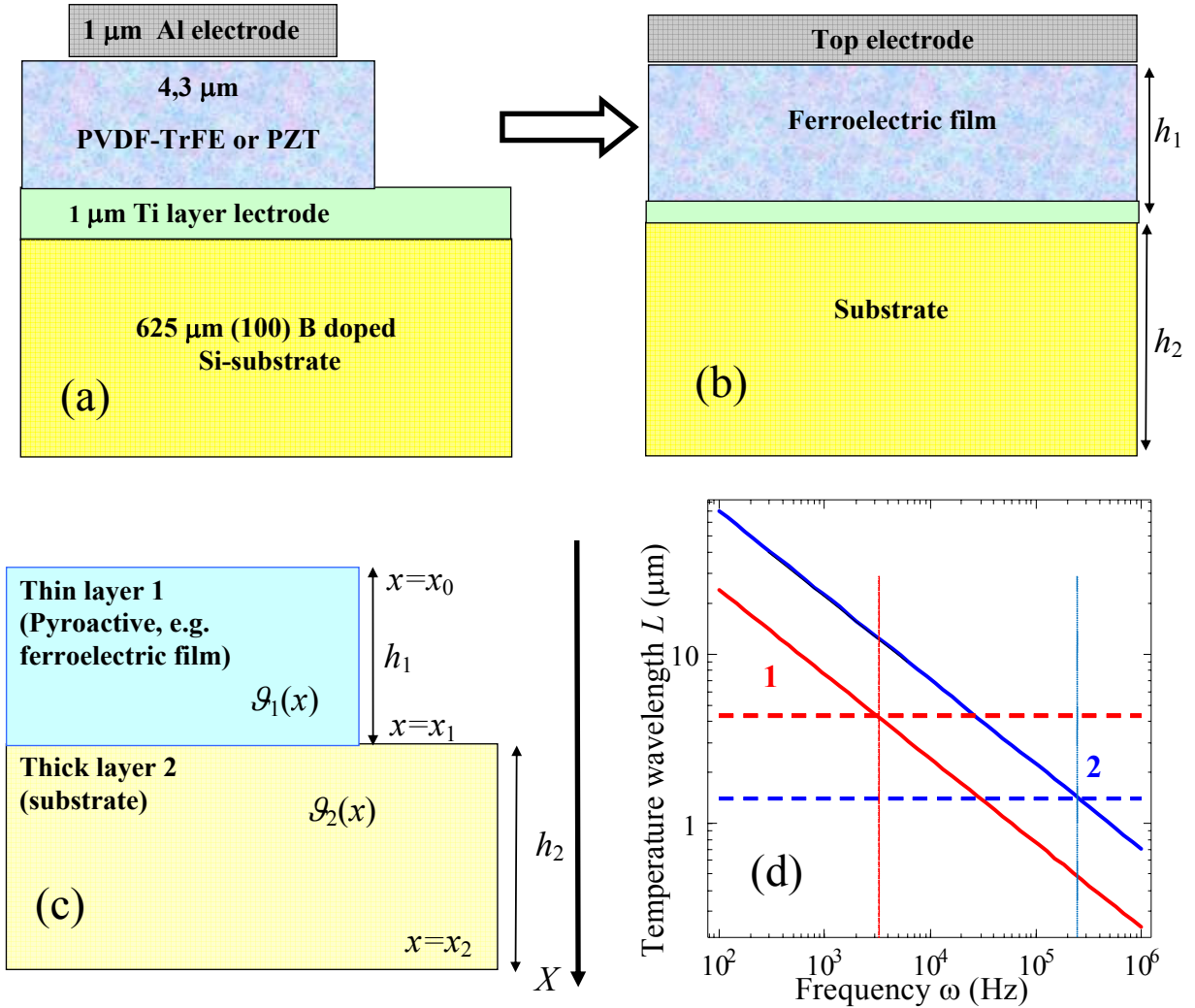

**Figure 1.** Schematic representation of a multi-layer pyroelectric element consisting of the top Al electrode, thin Ti-adhesive layer, ferroelectric P(VDF-TrFE) film, bottom Ti electrode and Si substrate **(a)**; simplified **(b)** and modeled **(c)** two-layer pyroactive structures. **(d)** Frequency dependence of the temperature wavelength (absolute value). Solid lines 1 and 2 correspond to the parameters of PVDF

and PZT respectively. Horizontal dashed line 1 corresponds to the PVDF film thickness used in the experiment, and the dashed line 2 is for the thickness of PZT film.

Heat flow across the *n*-layer system is described by a system of *n* linear thermal conductivity equations, and each layer is characterized by its own equation for the temperature variation inside the layer:

$$\frac{\partial}{\partial t}\vartheta_m(x,t) = \frac{k_m}{c_m}\frac{\partial^2}{\partial x^2}\vartheta_m(x,t) \tag{1}$$

Here $\vartheta_m$ is the temperature variation of the layer "*m*", $m=1, 2, \ldots n$, $c_m$ is a heat capacity and $k_m$ is a thermal conductivity. Since the relation between the heat flux and the temperature variation is given by the conventional expression, $j_m = -k_{Ti}\left.\frac{\partial \vartheta_m}{\partial x}\right|_{x=x_i}$, the boundary conditions to Eqs. (1) on the first "top" surface of the layer have the form of heat balance equation, $j_0(x_0) = \alpha F(t) - g_0\vartheta_1(x_1)$, where $\alpha$ is the absorption ability of surface, $F(t)$ is the incident heat flux, $t$ is the time, $g_0$ is the heat exchange coefficient between the surface and environment. The boundary conditions at other interlayer boundaries are the continuity of heat flux, $j_m(x_i) = j_{m+1}(x_i)$, and the equality of the layers' temperatures, $\left.\vartheta_m\right|_{x=x_i} = \left.\vartheta_{m+1}\right|_{x=x_i}$. The boundary condition on the "back" surface is $j_n(x_n) = g_n\vartheta_n(x_n)$, where $g_n$ is the coefficient of heat exchange of the surface with environment.

The incident heat flux $F(t)$ is a periodic function of time with modulation frequency $\omega$. Hence the periodic stationary solution of the Eqs.(1) has the form:

$$\vartheta_m(x,t) = \exp(i\omega t)[A_m \exp(iK_m x) + B_m \exp(-iK_m x)]. \tag{2}$$

The constants $A_m$ та $B_m$ should be calculated from the abovementioned boundary conditions. The dispersion relation that satisfies Eqs.(1) defines the constants $K_m$ as $K_m = \sqrt{-i\omega c_m/k_m}$. It appeared convenient for us to introduce the complex physical parameter that we called the "complex temperature wavelength",

$$L_m = \sqrt{a_m/(i\omega)}, \tag{3}$$

where the parameter $a_m = k_m/c_m$ is the thermal diffusivity. The characteristic depth of temperature wave attenuation is $|L_m| = \sqrt{a_m/\omega}$. After we represented the complex temperature wavelength in the form $L_m = \sqrt{k_m/i\omega c_m}$, the solution (2) acquires the following form $\vartheta_m(x,t) = \exp(i\omega t - x/\sqrt{2}|L_m|)[A_m \exp(ix/\sqrt{2}|L_m|) + B_m \exp(-ix/\sqrt{2}|L_m|)]$. As it follows from the expression, the "half" period of the thermal wave $\sqrt{2}\pi|L_m|$ is greater than the length of its attenuation. $\sqrt{2}|L_m|$, that corresponds to the concept of temperature wave as a strongly damped one [49].

The frequency dependence of the temperature wavelength's absolute value is shown by straight solid lines in **Figure 1d.** The line 1 corresponds to the parameters of ferroelectric P(VDF-TrFE), and the line 2 is for PZT. Horizontal dashed line 1 corresponds to the P(VDF-TrFE) film thickness used in the experiment, and the dashed line 2 is for the thickness of PZT film. The intersection points between the dashed and solid lines correspond to the frequencies 3 KHz and 200 KHz respectively. When the thickness of the film corresponds to the half-length of the temperature wave, $h_1 \sim \sqrt{2}\pi|L_m|$, the phenomenon of the interference (destructive or constructive) of the waves is possible due to reflection from the surface of the film [49].

The pyroelectric current $I_\pi = A_0 \left( \frac{\partial P_s}{\partial T} \frac{\partial \vartheta}{\partial t} \right)$ and voltage $U_\pi = \frac{A_0 \Delta P_s}{C} = A_0 \frac{\partial P_s}{\partial T} \frac{\Delta \vartheta}{C}$ are related with the temperature variation $\vartheta(x,t)$ in the following way:

$$I_\pi = \frac{\gamma}{h_1} A_0 \frac{\partial}{\partial t} \int_0^{h_1} \vartheta(x,t) dx, \quad U_\pi = \frac{\gamma}{\varepsilon_0 \varepsilon} \int_0^{h_1} \vartheta(x,t) dx. \quad (4)$$

Here $P_S$ is the spontaneous polarization of a ferroelectric film, $\gamma = \frac{\partial P_S}{\partial \vartheta}$ is its pyroelectric coefficient, and $C = \frac{\varepsilon_0 \varepsilon A_0}{h_1}$ is the pyroactive layer capacity. The temperature $T = T_0 + \vartheta(x,t)$. The average temperature variation of a pyroelectric film is:

$$\overline{\vartheta}(t) = \frac{1}{h_1} \int_0^{h_1} \vartheta(x,t) dx \quad (5)$$

From Eqs. (4) and (5) we obtain that

$$I_\pi = \gamma A_0 \frac{\partial \overline{\vartheta}}{\partial t} \qquad U_\pi = \frac{\gamma h_1}{\varepsilon \varepsilon_0} \overline{\vartheta} \quad (6)$$

To elucidate the frequency dependence of the experimentally measured pyroelectric voltage, let us consider a simplified equivalent circuit of pyroactive element shown in **Figure 2.** The complete circuit scheme can be found in refs [1-5].

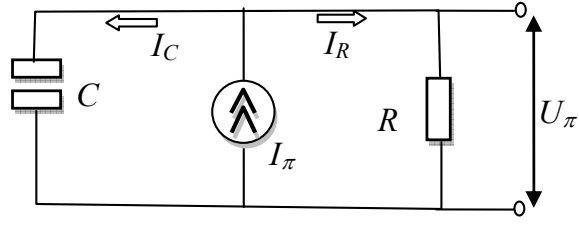

**Figure 2.** The equivalent circuit of pyroactive element. $R$ is the loading resistor, $C$ is the capacity of pyroactive layer, $I_\pi$ is the pyroelectric current generated by pyroactive layer as a current source, $I_C$ and $I_R$ are the currents in the $C$ and $R$ circuits respectively (adapted from ref [1]).

As one can see from **Figure 2** the following system of equations $I_R R = I_C Z_C$ and $I_R + I_C = I_\pi$ are valid for the currents in the circuit of pyro-active element. The the impedance and capacity of pyroactive layer are $Z_C = \dfrac{1}{i\omega C}$ and $C = \dfrac{\varepsilon\varepsilon_0 A_0}{h_1}$ correspondingly. Hence $I_C = \dfrac{R I_\pi}{R + Z_C}$, $I_R = \dfrac{Z_C I_\pi}{R + Z_C}$ and $U_\pi = I_R R = \dfrac{Z_C I_\pi}{R + Z_C} R = \dfrac{R I_\pi}{i\omega C R + 1}$. Then from expression (6) we obtain the dependence of pyro-voltage on the average temperature variation $\overline{\vartheta}$:

$$U_\pi = \frac{I_\pi R}{1 + i\omega RC} \equiv \frac{A_0 i\gamma\omega R}{1 + i\omega\tau_e}\overline{\vartheta}, \qquad (7)$$

where $\tau_e = RC$ is the electric relaxation time.

### III. MULTI-LAYER PYROACTIVE SYSTEMS: THE THEORY AND EXPERIMENT
### III.A. Analytical solution for a two-layer system

We studied experimentally a multi-layer system consisting of external (top) Al electrode, thin Ti-adhesive layer, P(VDF-TrFE) ferroelectric film, internal (bottom) Ti electrode, and Si-substrate [see **Figure 1a**]. Targeting to obtain closed-form analytical expressions and compare them with experiment, let us consider a simpler (somewhat idealized) two-layer system shown in **Fig. 1c**, since the temperature change in thin top and bottom metal electrodes with high thermal conductivity can be neglected for the approximate analysis.

The evident form of the Eqs. (1) and the boundary conditions for the two-layer system are presented in **Appendix A.** The solution of this system gives the following expression for the average temperature variation of the pyroactive ferroelectric film:

$$\overline{\vartheta}_1(\omega) = \frac{F_0 \alpha_1 L_1^2 \begin{pmatrix} k_2 L_2 \cosh(h_2/L_2)(k_1 \sinh(h_1/L_1) + g_2 L_1 (\cosh(h_1/L_1) - 1)) + \\ \sinh(h_2/L_2)\left(k_2^2 L_1 (\cosh(h_1/L_1) - 1) + g_2 k_1 L_2^2 \sinh\left(\dfrac{h_1}{L_1}\right)\right) \end{pmatrix}}{h_1 \begin{pmatrix} k_1 L_1 \cosh(h_1/L_1)((g_0 + g_2) k_2 L_2 \cosh(h_2/L_2) + (k_2^2 + g_0 g_2 L_2^2) \sinh(h_2/L_2)) + \\ + \sinh(h_1/L_1)(k_2 (k_1^2 + g_0 g_2 L_1^2) L_2 \cosh(h_2/L_2) + (g_0 k_2^2 L_1^2 + g_2 k_1^2 L_2^2) \sinh(h_2/L_2)) \end{pmatrix}} \quad (8)$$

In the expression (8) the complex temperature wavelengths $L_1$ and $L_2$ of the film and substrate are frequency dependent, namely $L_m = \sqrt{k_m/i\omega c_m}$ ($m = 1, 2$).

For low enough frequencies $\omega \ll \min_m \left\{\dfrac{k_m}{h_m^2 c_m}\right\}$, the layer thicknesses $h_m \ll |L_m|$, because $L_m \to \infty$. Approximate expression for the temperature variation $\overline{\vartheta}_1(\omega)$, obtained from the corresponding expansion of expression (7), is valid in the limiting case of low frequencies:

$$\overline{\vartheta}_1 \approx \frac{F_0 \alpha_1}{g_0}\left(1 - \frac{(h_1/2k_1) + (1/g_0)}{(h_1/k_1) + (h_2/k_2) + (1/g_0) + (1/g_2)}\right). \quad (9a)$$

Under the validity of strict inequalities $k_i \gg g_i |L_i|$ for intermediate frequencies, the ratio $h_m \sim |L_m|$. At that for intermediate frequencies $\max_m\left\{\dfrac{k_m}{h_m^2 c_m}\right\} \ll \omega \leq \max_m\left\{\dfrac{g_m^2}{k_m c_m}\right\}$ the frequency dependence of $\overline{\vartheta}_1(\omega)$ can be approximated as:

$$\overline{\vartheta}_1 \approx \frac{F_0 \alpha_1 L_1^2 (k_1 L_2 \cosh(h_2/L_2)\sinh(h_1/L_1) + k_2 L_1 \sinh(h_2/L_2)(\cosh(h_1/L_1) - 1))}{h_1 k_1 (k_1 L_2 \cosh(h_2/L_2)\sinh(h_1/L_1) + L_1 k_2 \cosh(h_1/L_1)\sinh(h_2/L_2))} \quad (9b)$$

When $h_2 \to \infty$ the expression (9b) can be simplified and acquires the form $\overline{\vartheta}_1 \approx \dfrac{F_0 \alpha_1 L_1^2 (k_1 L_2 \sinh(h_1/L_1) + k_2 L_1 (\cosh(h_1/L_1) - 1))}{h_1 k_1 (k_1 L_2 \sinh(h_1/L_1) + L_1 k_2 \cosh(h_1/L_1))}$.

For high enough frequencies the strict inequalities $k_m \gg g_m |L_m|$ and $h_m \gg |L_m|$ are valid simultaneously, because $L_m \to 0$. At that $\omega \gg \max_m\left\{\dfrac{k_m}{h_m^2 c_m}, \dfrac{g_m^2}{k_m c_m}\right\}$ and the simple approximation for temperature variation is valid

$$\overline{\vartheta}_1 \approx \frac{F_0 \alpha_1}{\omega c_1 h_1}, \quad (9c)$$

As it follows from expressions (9) that the temperature variation depends on the heat exchange constants $g_0$ and $g_2$ only at low frequencies, at that it tends to a certain constant value. For intermediate frequencies the frequency dependence $\overline{\vartheta}_1(\omega)$ is much more complicated. At high frequencies, the temperature variation decreases with increasing frequency as $1/\omega$.

According to the expressions (6) and (7) pyro-current and pyro-voltage are complex functions of $\overline{\vartheta}_1(\omega)$, in particular, $I_\pi(\omega) = i\omega\gamma A_0 \overline{\vartheta}_1$ and $U_\pi(\omega) = \dfrac{A_0 i\gamma\omega R}{1+i\omega\tau_e}\overline{\vartheta}_1$. The amplitude and phase of the pyro-current and pyro-voltage frequency spectra will be analyzed below and compared to the ones obtained experimentally in refs [46-48].

The frequency spectra of the amplitude and phase of $\overline{\vartheta}_1(f)$, $I_\pi(f)$ and $U_\pi(f)$ are presented in **Figures 3**. The linear frequency $f = \omega/2\pi$. Different curves 1-4 correspond to the different values of heat exchange constants $g_0$ and $g_2$, changing from small ~1 to large ~$10^3$ values (here and below in W/m$^2$K). The dependences are normalized on the value $\gamma F_0 \alpha_1$.

At low frequencies $\overline{\vartheta}_1(f)$ is frequency independent and its value decreases with $g_0$ increasing (**region I**). At that $I_\pi(f)$ and $U_\pi(f)$ amplitudes increase with frequency increasing. In accordance with Eqs.(9), $I_\pi(f)$ and $U_\pi(f)$ depend on the constants $g_0$ and $g_2$ only at low frequencies $f < 10$ Hz.

For intermediate frequencies 10 Hz $< f <$ 10 kHz the amplitudes of $\overline{\vartheta}_1(f)$, $I_\pi(f)$ and $U_\pi(f)$ are frequency dependent and have a complicate shape including "plateau" region limited by the regions of $1/f$-decrease for $\overline{\vartheta}_1(f)$ and $f$-increase for $I_\pi(f)$ and $U_\pi(f)$ correspondingly (**region II**). For high frequencies $f >> 10$ kHz the amplitudes of $\overline{\vartheta}_1(f)$ and $U_\pi(f)$ decrease with increasing frequency as $1/f$, while the amplitude of $I_\pi(f)$ tends to a constant value (**region III**).

**Figure 3a** shows the frequency spectra of the amplitude $\overline{\vartheta}_1(f)/F_0\alpha_1$ at $g_0 = g_2 = 1$, 10, $10^2$, $10^3$ W/m$^2$K (see curves 1-4). At $g_0 = g_2 = 1$ W/m$^2$K (curve 1) the temperature variation decreases as $1/f$ under the frequency increasing. The curves 2-4 are similar to the curve 1, however the low frequency plateau region extends to higher frequencies under $g_0$ and $g_2$ increase.

The phase of $\overline{\vartheta}_1(f)$, shown in **Figure 3b**, has a pronounced minima at the boundaries of the regions I, II and III, where it approaches $-90°$. The values of the temperature wavelength, estimated for the frequencies 100 Hz and 10 kHz, are approximately equal to 650 μm and 4 μm, that correspond to the boundaries between the regions I and II, and II and III (see the vertical lines in **Fig. 3b**). The values are close to substrate thickness 625 μm and film thickness 4.3 μm (see **Fig. 1a**). So the features of the $\overline{\vartheta}_1(f)$ frequency spectra strongly correlate with the structure of the investigated two-layer system.

The frequency spectra of the normalized amplitude and phase of the pyroelectric current $I_\pi(f)$ and voltage $U_\pi(f)$ are shown in **Figure 3c-f**. The frequency spectra of $I_\pi(f)$ amplitude and phase calculated at $g_0 = g_2 = (1, 10, 10^2, 10^3)$ W/m$^2$K are shown by curves 1-4 in **Figures 3c** and **3d**. The amplitude of $I_\pi(f)$ is constant in the low-frequency region I at $g_0 = g_2 = 1$ W/m$^2$K (curve 1), then it increases proportionally to $f$, reaches a weakly pronounced maximum at the boundary between the regions II and III and becomes constant in the high frequency region III. The curves 2-4 are similar to the curve 1 in the regions II and III, but they are different in the region I, where the cut-off frequency depends on the $g_0 = g_2$ value. The phase of $I_\pi(f)$ differs from the $\overline{\vartheta}_1(f)$ phase on the factor $\pi/2$ (see **Fig. 3b**), because $I_\pi(f) \sim if\overline{\vartheta}_1(f)$ accordingly to the Eq. (6).

The overall view and features of the curve 2 in **Figure 3c** are similar to those calculated by Mahrane et al. [29] and Simonne et al. [30] for P(VDF-TrFE)-on-Si structure (see **Fig. 2** in these papers). Some differences in the frequency positions of the features (except for the boundary between the regions I and II) are connected with different structure type and configuration. Namely a three-layer structure with 10 μm thick thermal insulator and film and substrate thicknesses 10 μm and 250 μm was considered in [29, 30], while we consider a two-layer system.

**Figures 3e** and **3f** show the frequency spectra of the $U_\pi(f)$ amplitude and phase calculated at $g_0 = g_2 = 1, 10, 10^2, 10^3$ W/m²K (curves 1-4). The amplitude of $U_\pi(f)$ is constant in the low-frequency region I at $g_0 = g_2 = 1$ W/m²K (curve 1), then it increases proportionally to $f$ in the region II reaching a well pronounced maximum at the boundary of regions II and III, and then it decreases proportionally to $1/f$ in the high frequency region III. At $g_0 = g_2 = 10^3$ W/m²K (curve 4) $U_\pi(f)$ increases proportionally to $f$ in the region I, has an inflexion point at the region I-II boundary and then behaves as curve 1. The curves 2 and 3 are similar to the curve 4, but with other cut-off frequency positions in the region I. The phase of $U_\pi(f)$ does not reproduce the phase of $\overline{\vartheta}_1(f)$ (see **Fig. 3b**) because they differ on the frequency factor $\dfrac{A_0 i\gamma f R}{1 + 2\pi i f \tau_e}$ in accordance with Eq.(7).

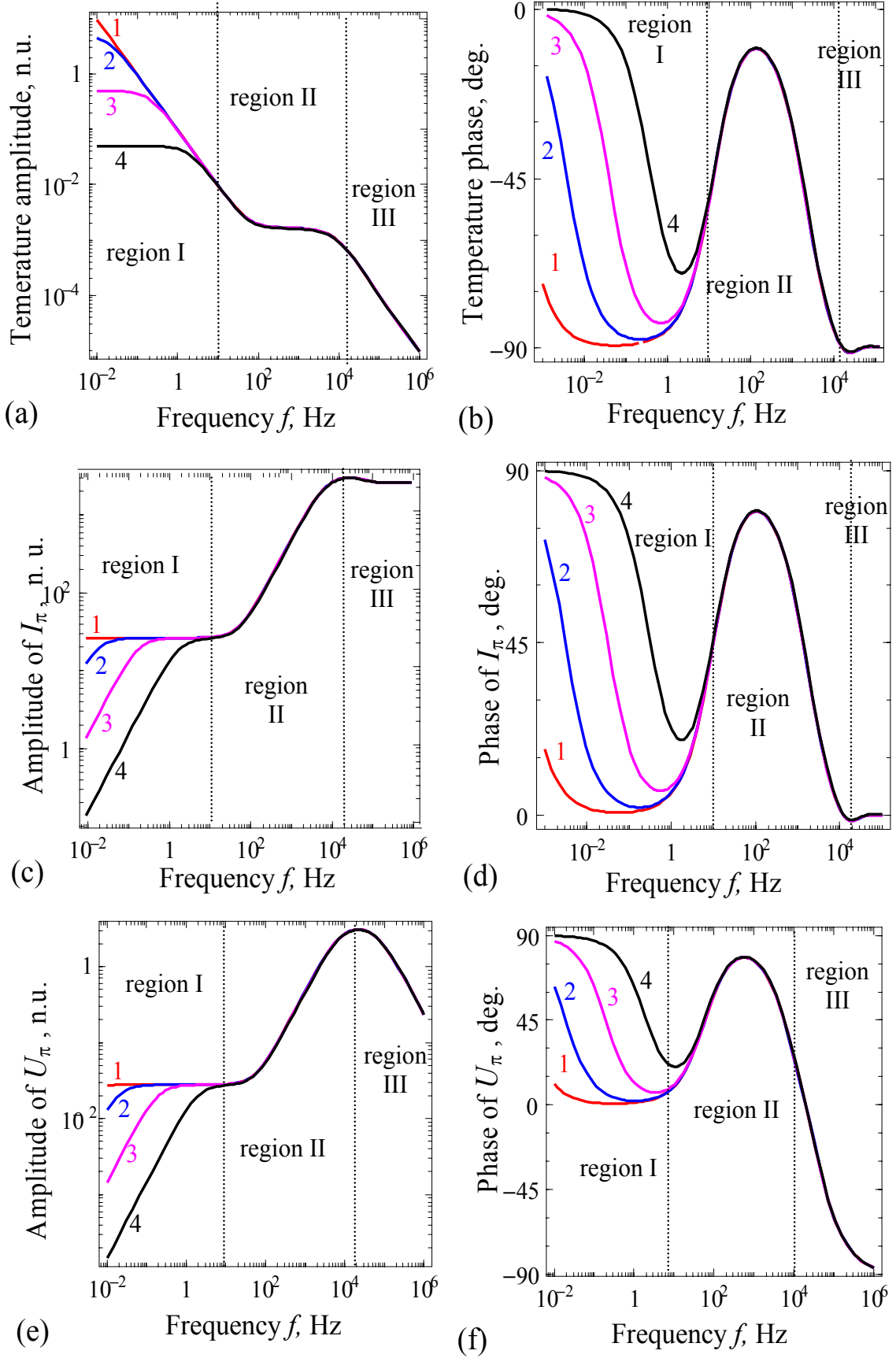

**Figure 3.** Frequency spectra of amplitude **(a, c, e)** and phase **(b, d, f)** of the average temperature variation $\overline{\vartheta}_1$ (a, b), pyroelectric current $I_\pi(f)$ (c, d) and voltage $U_\pi(f)$ (e, f) of ferroelectric P(VDF-

TrFE) film on Si-substrate. Different curves are calculated for different values of heat exchange constants $g_0 = g_2 = (1; 10; 10^2; 10^3)$ W/m²K (curves 1-4). The film thickness $h_1 = 4.3$ μm, substrate thickness $h_2 = 625$ μm, thermal conductivities $k_1 = 0.15$ W/Km and $k_2 = 150$ W/Km, heat capacities $c_1 = 2.1 \cdot 10^6$ J/m³K and $c_2 = 1.6 \cdot 10^6$ J/m³K, electrode area $A_0 = (1.7 \cdot 10^{-3})^2$ m²; dielectric permittivity $\varepsilon = 20$. The dependences are normalized on the factor $\gamma F_0 \alpha_1$. Abbreviation "n. u." signifies normalized units.

### III.B. Size effect of pyroelectric response

The frequency spectra of the amplitude and phase of the temperature variation $\overline{\vartheta}_1$, pyroelectric current $I_\pi$ and voltage $U_\pi$ were calculated for different values of the pyroelectric layer thickness $h_1$. Results are presented in **Figure 4** for $h_1$ = 0.1, 1, 10 and 100 μm (curves 1-4).

Note that we have not considered very thin ferroelectric films of a thickness less then 100 nm with inherent "internal" size effects, including thickness dependence of the spontaneous polarization and pyroelectric coefficient, caused by the influence of so-called extrapolation length [50, 51, 52]. In particular the distribution of polarization and pyroelectric coefficient becomes substantially inhomogeneous at small extrapolation lengths, and therefore they cannot be considered as spatially uniform in the calculations. Thus the frequency spectra shown in **Figure 4** correspond to "external" size effect of the temperature variation and pyroelectric response characteristics. Note the strong dependence of the peculiarities of $\overline{\vartheta}_1(f)$, $I_\pi(f)$ and $U_\pi(f)$ maxima on the ferroelectric film thickness $h_1$.

**Figures 5a** and **5b** show the dependences of $|U_\pi(f)|$ maximum value and its frequency position on the ferroelectric film thickness $h_1$. Since these dependences are non-monotonic, it makes sense to search for the optimal thickness $h_{opt}$ and optimal frequency $f_{opt}$ in the specified "working" range of frequencies that determines the maximal sensitivity of pyroactive element. As it follows from **Figs 5**, $h_{opt}$ is equal to 4 μm and $f_{opt}$ is equal to 2.5 kHz for the considered two-layer system.

The dependences of pyro-current $I_\pi$ and pyro-voltage $U_\pi$ on the frequency $f$ and on the ferroelectric film thickness $h_1$ are illustrated by the corresponding $f$-$h_1$-surfaces in **Figure 6**.

The pyro-current $f$-$h_1$-surface (**Fig. 6a**) has a crest-like maximum curved in accordance with $L_m(\omega)$ dependence (see expression (3)). It is worth to note practical equality of frequency positions ($\approx 10^3$ Hz) of the crest in **Fig. 6a** and normalized current diffuse maximum in [29, 30] for a 10 μm thick P(VDF-TrFE) film.

The pyro-voltage $f$-$h_1$-surface (**Fig. 6b**) has a ridge-like maximum that position weakly shifts to low frequencies and diffuseness increases with film thickness increasing. The acme of the $f$-$h_1$-surface is positioned at $3.5 \cdot 10^3$ Hz and 4.5 μm.

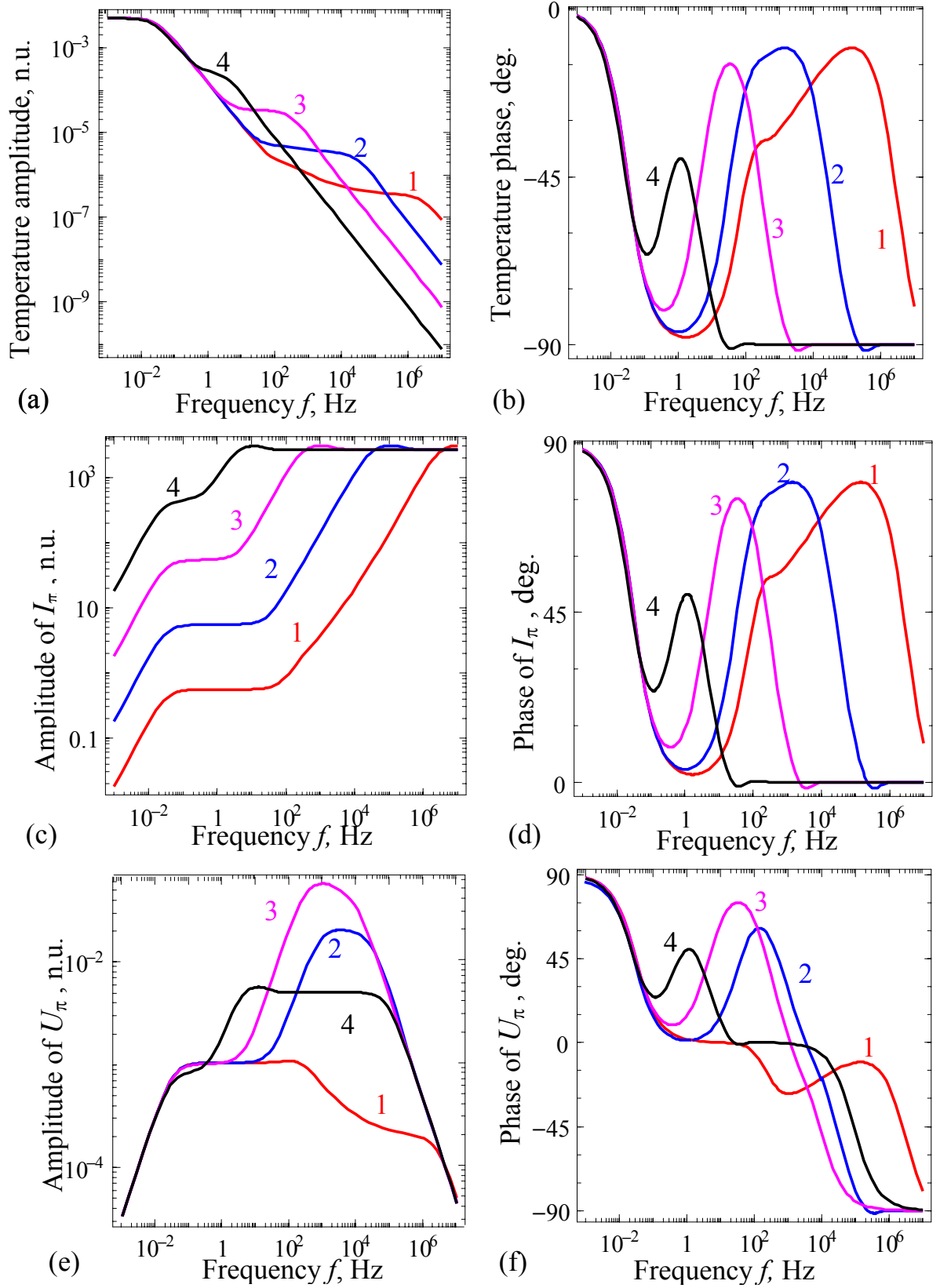

**Figure 4.** Frequency spectrum of the average temperature variation $\bar{\vartheta}_1(f)$ **(a, b)**, pyro-current $I_\pi(f)$ **(c, d)** and pyro-voltage $U_\pi(f)$ **(e, f)** calculated for different thickness $h_1$ of ferroelectric film, 0.1; 1;

10; 100 μm (curves 1-4), substrate thickness $h_2 = 625$ μm and heat exchange constants $g_0 = g_2 = 10^2$ W/m$^2$K. Other parameters are the same as in **Figure 3**.

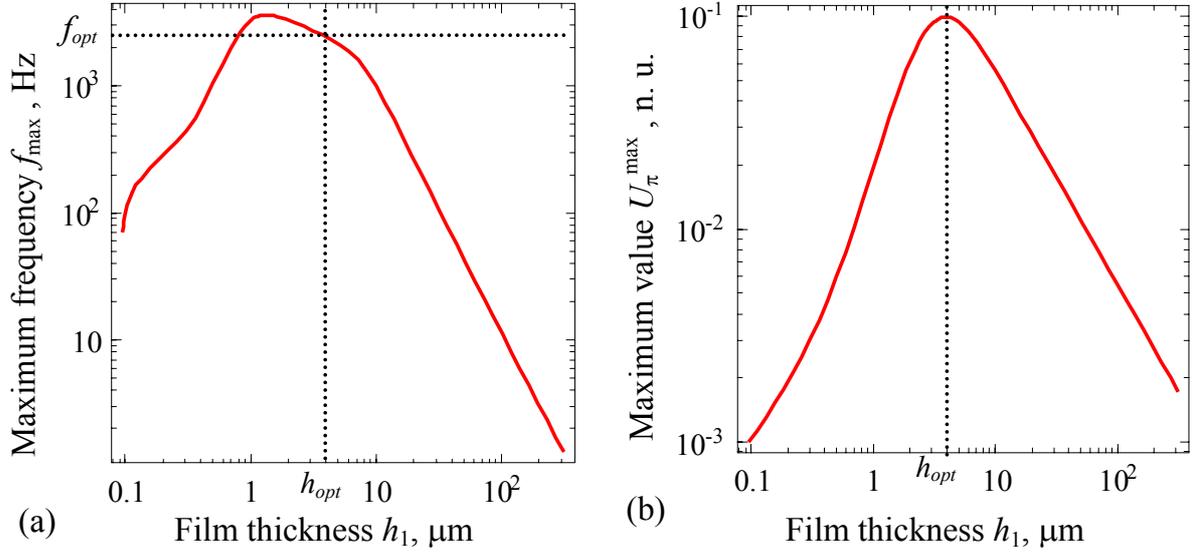

**Figure 5.** Dependence of the $U_\pi$ maximum frequency position $f_{max}$ **(a)** and absolute value **(b)** on the ferroelectric film thickness $h_1$, calculated for the same parameters as in **Figure 4**.

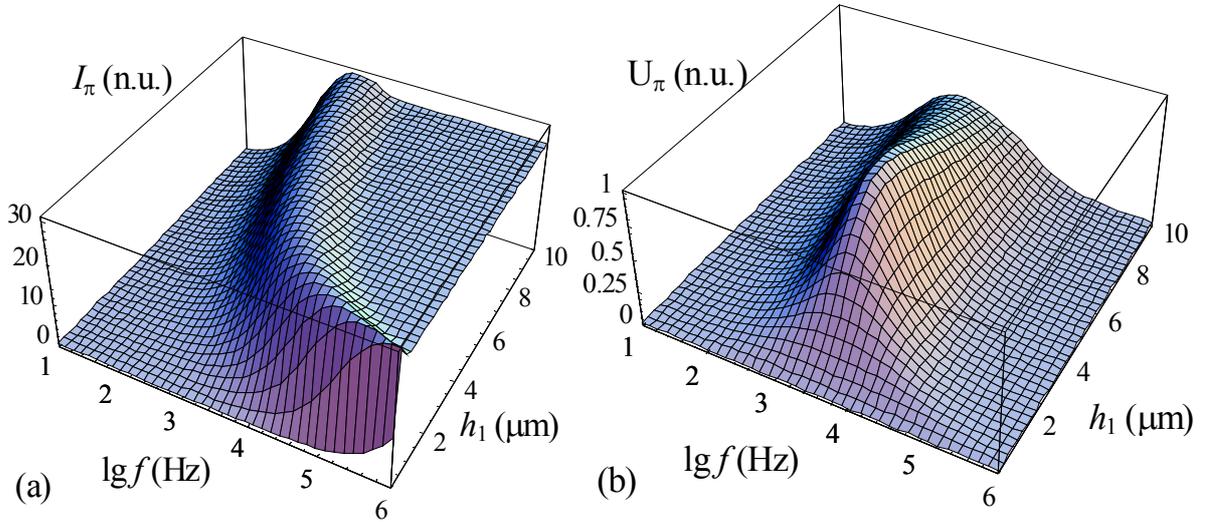

**Figure 6.** The dependences of pyro-current $I_\pi$ **(a)** and pyro-voltage $U_\pi$ **(b)** on the frequency $f$ and ferroelectric film thickness $h_1$, calculated for the same parameters as in **Figure 4.**

Allowing for the existence of the size effect, the analysis of analytical expressions (8)-(9) can give the recommendations for the optimization of pyroactive elements performances depending on the material parameters of pyroactive layer (heat conductivity $k_1$ and capacity $c_1$, dielectric permittivity ε), its thickness $h_1$, material parameters of the substrate ($k_2$, $c_2$ values) and its thickness $h_2$, and heat exchange constants on the layers boundaries ($g_0$, $g_2$), which are strongly dependent on the electrode material.

### III.C. Comparison with experiment

Schematic representation of multi-layer pyroelectric elements consisting of the top Al electrode, thin Ti-adhesive layer, ferroelectric (poly-vinylidene fluoride -trifluoroethylene copolymer P(VDF-TrFE) or lead zirconate-titanate 40/60, PZT) film, bottom Ti electrode and Si substrate, which were studied experimentally [46-48], is shown in the **Figure 1a.** The simplified model of such element is shown in the **Figure 1c**. The frequency spectra of amplitude and phase of pyroelectric voltage generated by pyroelectric element Al/P(VDF-TrFE)/Ti/Si are presented in **Figure 7**. Symbols are the spectra measured by S. Bravina et al [45, 46]; solid and dotted curves are calculated for relatively low and relatively high values of the loading resistance $R_L$.

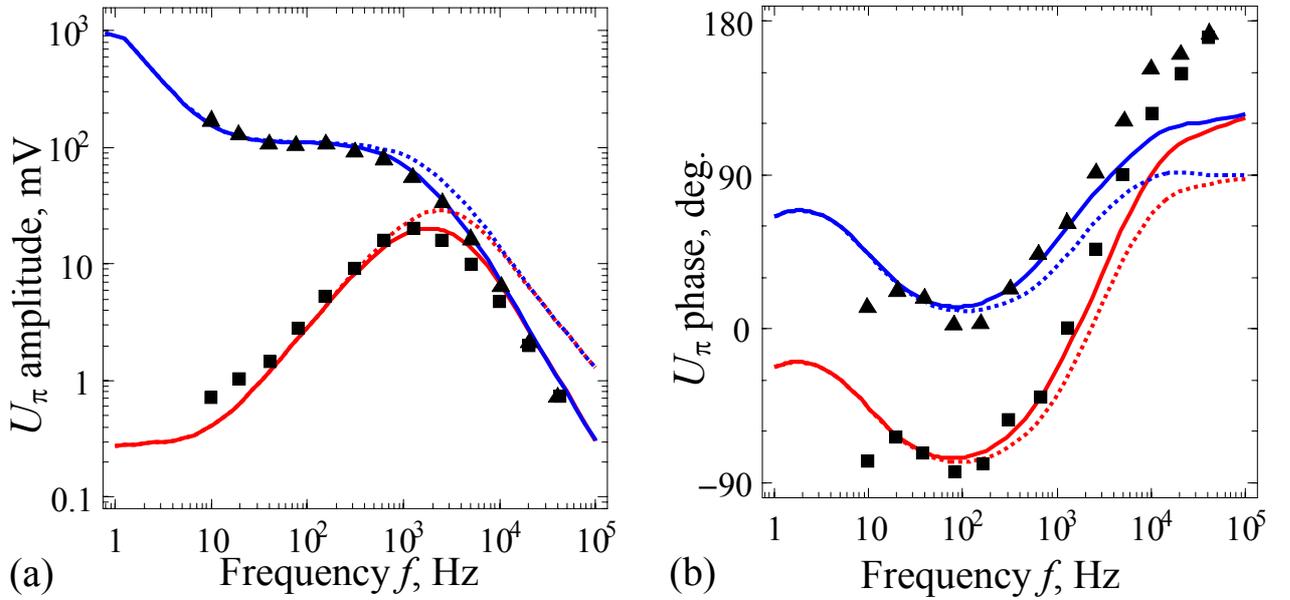

**Figure 7.** Frequency spectra of the pyroelectric voltage $U_\pi(f)$ amplitude (a) and phase (b) measured experimentally (symbols) and calculated theoretically (solid and dotted curves) for Al/P(VDF-TrFE)/Ti/Si structure at different values of loading resistance $R_L = 10^5$ Ω (squares), $R_L = 10^{10}$ Ω (triangles). The thickness of the P(VDF-TrFE) film $h_1 = 4.3$ μm, the thickness of silicon substrate $h_2 = 625$ μm. Other parameters are the same as in **Figure 3**.

The solid curves are calculated taking into account the thermal influence of the top 1 μm-thick Al electrode and the dotted curves are calculated without it. By taking into consideration the heat transfer delay in the top electrode we reach significantly better agreement between the theoretically calculated and experimentally measured amplitudes of pyroelectric voltage at high frequencies.

Amplitude of pyroelectric voltage $U_\pi(f)$ for high $R_L$ decreases non-monotonically with increasing of the modulation frequency $f$. In contrast, $U_\pi(f)$ increases for low $R_L$ at first, then it has a maximum at a frequency about 2 kHz, and then it decreases in accordance with the law $1/f$. The amplitudes of low-frequency pyroelectric voltage are larger at high $R_L$ than those at low $R_L$, but they all become the same at high frequencies. The result is consistent with the expression (9), because the pyroelectric voltage is not dependent on $R_L$ at high frequencies, namely $U_\pi \underset{\omega\to\infty}{\to} \dfrac{A_0 \gamma}{C}\overline{\vartheta}$.

As for the frequency spectrum of $U_\pi$-phase, it has a similar behavior for low and high resistances $R_L$. At first it decreases with $f$ increase, then falls to a minimum at the frequency about 100 Hz, and then increases and tends to $180^\circ$ under further frequency increasing. The phase calculated theoretically saturates at frequencies higher than 10 kHz (see dotted curves in **Figure 7b**). From $\lambda_T$ equality point of view, Al layer of 1 μm thick is equivalent to PVDF layer of 0.03 μm thick. So, the discrepancy between the experimentally measured and theoretically calculated phases at high frequencies can be associated with sub-surface inhomogeneties inherent to PVDF films [53, 54, 55], which are ignored in the simplified model (compare solid and dashed curves in **Fig. 7b**). Despite the warning, we demonstrated that the proposed theoretical model can quantitatively describe experimental results in the actual frequency range.

## IV. CONCLUSION

In the framework of the heat conduction theory we have developed a theoretical model for the description of pyroelectric response frequency spectra in multilayer system "top electrode/pyroactive ferroelectric film/bottom electrode/substrate", which are basic elements of modern multifunctional pyroelectric detectors of irradiation and transducers.

We solved analytically the system of the thermal conductivity equations with the mixed-type boundary conditions of the third kind at the interfaces of the multilayer system. The periodic change of pyroactive element temperature was caused by an incident modulated heat flux. Analytical expressions for the frequency spectra of temperature variation and pyroelectric response characteristics (such as pyroelectric current and voltage) were derived and graphically analyzed.

The dependence of the pyroelectric response frequency spectra on the ferroelectric film thickness has been derived and the existence of pronounced extrinsic size effect has been revealed. In particular, the thickness dependence of the frequency position of the pyroelectric response features has been found, at that the value and frequency position of the pyroelectric voltage maximum differ in more than one order of magnitude under less than tenfold changing of the ferroelectric film thickness.

The dependences of the frequency position and height of the pyroelectric voltage maximum (actual sensitivity of the system) on the film thickness were calculated and analyzed. As an example the optimal thickness of the pyroelectric layer (≈ 4 μm) and optimal operating frequency (≈ 2.5 kHz) for the system Al/P(VDF-TrFE)/Ti/Si were determined. The proposed algorithm can be one of the ways to increase the sensitivity and to improve the working performances of multifunctional pyroelectric transducers.

Basing on the obtained results, one can give the recommendations for optimization of the pyroelectric detectors sensitive elements by using the relation between the extrinsic size effect and material parameters of pyroactive film and substrate, such as heat conductivity and capacity, dielectric permittivity and thickness, as well as heat exchange conditions at the interfaces of the pyroactive multilayer.

**Authors' contribution.** E.A.E. performed analytical and numerical calculations jointly with M.Y.Y. N.V.M. obtained the experimental results and wrote the introductive part of the paper. A.N.M. generated the idea of the research, stated the problem and wrote the original part of the paper. All co-authors discuss the obtained results and worked on the improvement of the manuscript. E.A.E., N.V.M. and A.N.M. acknowledge National Academy of Sciences of Ukraine (grant 07-06-15).

# APPENDIX A

# ANALYTICAL SOLUTION OF THE THERMAL CONDUCTIONVITY EQUATIONS

,The system of two equations for the case of two layers (according to the number of the layers) can be written as follows:

$$\begin{cases} \dfrac{\partial}{\partial t}\vartheta_1(x,t) = \dfrac{k_1}{c_1}\dfrac{\partial^2}{\partial x^2}\vartheta_1(x,t) \\ \dfrac{\partial}{\partial t}\vartheta_2(x,t) = \dfrac{k_2}{c_2}\dfrac{\partial^2}{\partial x^2}\vartheta_2(x,t) \end{cases} \quad (A.1)$$

The boundary conditions are:

$$\begin{cases} -k_1\dfrac{\partial\vartheta_1}{\partial x}\bigg|_{x=x_0} = \alpha_1 F(t) - g_0\vartheta_1\big|_{x=x_0} \\ -k_1\dfrac{\partial\vartheta_1}{\partial x}\bigg|_{x=h_1} = -k_2\dfrac{\partial\vartheta_2}{\partial x}\bigg|_{x=h_1} \\ \vartheta_1\big|_{x=h_1} = \vartheta_2\big|_{x=h_1} \\ -k_2\dfrac{\partial\vartheta_2}{\partial x}\bigg|_{x=h_1+h_2} = g_2\vartheta_2\big|_{x=h_1+h_2} \end{cases} \quad (A.2)$$

The general solution of the system (A.1) has the form:

$$\begin{cases} \vartheta_2(x,t) = A_2\exp(i\omega t)(\exp[(x-h_1)/L_2] + B_2\exp[-(x-h_1)/L_2]) \\ \vartheta_1(x,t) = A_1\exp(i\omega t)(\exp[(x-x_0)/L_1] + B_1\exp[-(x-x_0)/L_1]) \end{cases} \quad (A.3)$$

Substituting expressions (A.3) into the boundary conditions (A.2), we could obtain a system of algebraic equations for the determination of unknown constants $A_i$ and $B_i$:

$$\begin{cases} -k_1(A_1/L_1 + B_1/L_1) = \alpha_1 F(t) - g_0(A_1 + B_1), \\ (k_1/L_1)(A_1\exp((h_1-x_0)/L_1) + B_1\exp[-(h_1-x_0)/L_1]) = (k_2/L_2)(A_2+B_2), \\ A_1\exp[(x-x_0)/L_1] + B_1\exp[-(x-x_0)/L_1] = [A_2+B_2]/L_2, \\ -(k_2/L_2)[A_2\exp[h_2/L_2] + B_2\exp[-h_2/L_2]] = g_2 A_2\exp[h_2/L_2] + B_2\exp[-h_2/L_2]. \end{cases} \quad (A.4)$$

The amplitude of average temperature variation was found from this system:

$$\overline{\vartheta}_1(\omega) = \dfrac{F_0\alpha_1 L_1^2 \begin{pmatrix} k_2 L_2 \cosh\left(\dfrac{h_2}{L_2}\right)\left(k_1\sinh\left(\dfrac{h_1}{L_1}\right) + g_2 L_1\left(\cosh\left(\dfrac{h_1}{L_1}\right) - 1\right)\right) + \\ \sinh\left(\dfrac{h_2}{L_2}\right)\left(k_1^2 L_1\left(\cosh\left(\dfrac{h_1}{L_1}\right) - 1\right) + g_2 k_1 L_2^2 \sinh\left(\dfrac{h_1}{L_1}\right)\right) \end{pmatrix}}{h_1\begin{pmatrix} k_1 L_1 \cosh\left(\dfrac{h_1}{L_1}\right)\left((g_0+g_2)k_2 L_2\cosh\left(\dfrac{h_2}{L_2}\right) + (k_2^2 + g_0 g_2 L_2^2)\sinh\left(\dfrac{h_2}{L_2}\right)\right) + \\ +\sinh\left(\dfrac{h_1}{L_1}\right)\left(k_2(k_1^2 + g_0 g_2 L_1^2)L_2\cosh\left(\dfrac{h_2}{L_2}\right) + (g_0 k_2^2 L_1^2 + g_2 k_1^2 L_2^2)\sinh\left(\dfrac{h_2}{L_2}\right)\right) \end{pmatrix}} \quad (A.5)$$

**Figure A.1** shows how that temperature variation $\bar{\vartheta}_1$ and the pyro-current $I_\pi$ generated changes under the ferroelectric film thickness increase.

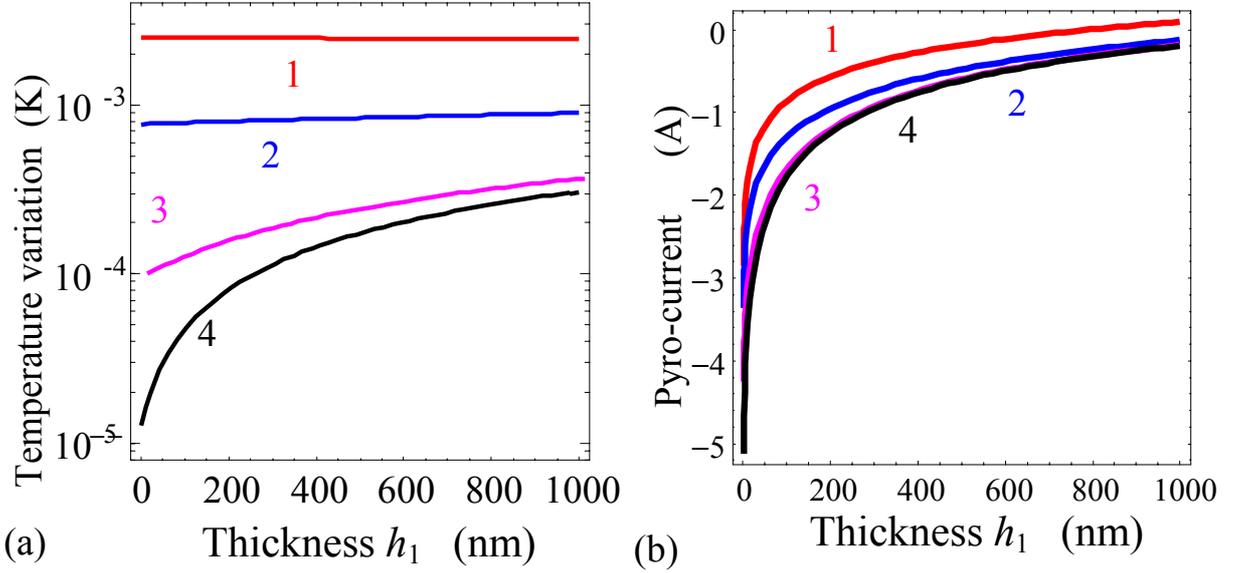

**Figure A1.** Thickness dependence of temperature variation $\bar{\vartheta}_1$ (a) and current $I_\pi$ (b), calculated for the two-layer system. The values of the parameters $g_2 = 10^2$, $g_2 = 10^3$; $g_2 = 10^4$, $g_2 = 10^5$ correspond to the curves colors (red (1), blue (2), purple (3), black (4 )) respectively. $g_0 = 300$, $g_3 = 50$. frequency $\omega = 10^{-4}$ Hz.